\documentclass[fleqn,usenatbib]{mnras}

\usepackage{newtxtext,newtxmath}

\usepackage[T1]{fontenc}

\DeclareRobustCommand{\VAN}[3]{#2}
\let\VANthebibliography\thebibliography
\def\thebibliography{\DeclareRobustCommand{\VAN}[3]{##3}\VANthebibliography}

\usepackage{graphicx}	
\usepackage{amsmath}	
\usepackage{academicons}
\usepackage{xcolor}
\usepackage{fontawesome5}
\usepackage{hyperref}

\newcommand{\Mp}{M_{\rm p}}

\definecolor{orcidlogocol}{rgb}{0.65, 0.807, 0.223}

\newcommand{\orcid}[1]{$\,$\href{https://orcid.org/#1}{\textcolor{orcidlogocol}{\faOrcid}}}

\title[PBH wide binary constraints]{Modelling uncertainties in wide binary constraints on primordial black holes}

\author[E. Tyler et al.]{
Emily Tyler,$^{1}$\orcid{0000-0002-3228-5499}
Anne M. Green,$^{1}$\thanks{E-mail: anne.green@nottingham.ac.uk}\orcid{0000-0002-7135-1671}
and Simon P. Goodwin$^{2}$\orcid{0000-0001-6396-581X}
\\
$^{1}$ School of Physics and Astronomy, University of Nottingham, Nottingham, NG7 2RD, UK\\
$^{2}$ Department of Physics and Astronomy, University of Sheffield, Sheffield, S3 7RH, UK\\
}

\date{Accepted XXX. Received YYY; in original form ZZZ}

\pubyear{2022}

\begin{document}
\label{firstpage}
\pagerange{\pageref{firstpage}--\pageref{lastpage}}
\maketitle

\begin{abstract}
Dark matter in the form of compact objects with mass $M_{\rm co} \gtrsim 10 M_{\odot}$ can be constrained by its dynamical effects on wide binary stars. Motivated by the recent interest in Primordial Black Hole dark matter, we revisit the theoretical modelling involved in these constraints. We improve on previous studies in several ways. Specifically, we i) implement a physically motivated model for the initial wide-binary semi-major axis distribution, ii) include unbound binaries, and iii) take into account the uncertainty in the relationship between semi-major axis and observed angular separation. These effects all tend to increase the predicted number of wide binaries (for a given compact object population). Therefore the constraints on the halo fraction in compact objects, $f_{\rm co}$, are significantly weakened. For the 
wide binary sample used in the most recent calculation of the constraints, we find the fraction of halo dark matter in compact objects is $f_{\rm co} <  1$ for $M_{\rm co} \approx 300 \, M_{\odot}$, tightening with increasing 
$M_{\rm co}$ to $f_{\rm co} < 0.26$ for  $M_{\rm co} \gtrsim 1000 \, M_{\odot}$. 
\end{abstract}




\section{Introduction}
There is strong evidence from cosmological and astronomical observations that $\approx 85 \%$ of the matter in the Universe is in the form of cold, nonbaryonic dark matter (DM), see e.g.~\citet{Bertone:2004pz} for a review. Traditionally the most popular dark matter candidates have been new elementary particles, such as Weakly Interacting Massive Particles or axions. However, the discovery of gravitational waves from mergers of tens of Solar mass black holes by LIGO-Virgo~\citep{LIGOScientific:2016aoc} has led to a surge of interest in Primordial Black Holes (PBHs) as a dark matter candidate~\citep{Bird:2016dcv,Sasaki:2016jop,Carr:2016drx}. PBHs are black holes that may form in the early Universe, for instance from the collapse of large density perturbations~\citep{zn,Hawking:1971ei}.

There are various constraints on the abundance of PBHs with mass $M_{\rm PBH} \gtrsim 1 M_{\odot}$ from gravitational microlensing~\citep{Diego:2017drh,Zumalacarregui:2017qqd,2022arXiv220213819B,2022ApJ...929L..17E}, gravitational waves from 
mergers of binaries~\citep{Sasaki:2016jop,Ali-Haimoud:2017rtz}, their dynamical effects on stars in wide binaries~\citep{Yoo:2003fr,Quinn:2009zg,mr}, and in dwarf galaxies~\citep{Brandt:2016aco}, and the radiation emitted due to accretion of gas onto PBHs~\citep{Ricotti:2007au,Gaggero:2016dpq}. For reviews, with extensive reference lists, see e.g.~\citet{Carr:2020xqk,Green:2020jor}.  The increased interest in PBH DM motivates a careful reanalysis of these constraints. For instance, the constraints from the temperature anisotropies in the Cosmic Microwave Background, due to the effects of PBHs on the recombination history of the Universe, have been found to be significantly weaker than previously thought~\citep{Ali-Haimoud:2016mbv,Poulin:2017bwe}.

In this paper we focus on the constraints on multi-Solar mass compact objects in the halo of the Milky Way (MW) from their dynamical effects on wide binary stars. While this is motivated by the recent interest in PBHs as a dark matter candidate, these constraints apply to any compact object DM. Close encounters between binary stars and massive compact objects increase the energies and semi-major axes of the binaries, and potentially disrupt some of the binaries. Observations of the semi-major axis distribution of wide binaries in the MW can therefore potentially constrain the abundance of compact objects. For perturbers with mass $M_{\rm p} \gtrsim 10^{3} M_{\odot}$ the closest encounter dominates, while for lighter perturbers it is necessary to take into account the cumulative, diffusive, effects of multiple interactions~\citep{BHT,b+t}.

\citet{BHT} used wide binaries in the Milky Way disk to constrain the fraction of the local mass density in compact objects. \citet{Yoo:2003fr} then used a sample of 90 wide halo binaries compiled by \citet{CG} to constrain the fraction of the MW halo in compact objects. They found that compact objects with mass $M_{\rm co} > 43 \, M_{\odot}$ could not make up all of the halo, and objects with mass $M_{\rm co} \gtrsim 10^{3} M_{\odot}$ were constrained to make up less than $20\%$ of the halo, at $95\%$ confidence. 

\citet{Quinn:2009zg}  highlighted that these constraints are very sensitive to the widest binaries. They carried out radial velocity measurements of four of the widest binaries in the \citet{CG} sample, and found that the second widest binary was in fact not a binary, as the two stars have significantly different radial velocities. Without this spurious binary, the mass above which compact objects were excluded from making up all of the halo increased to $M_{\rm co} \sim 500 \, M_{\odot}$. The radial velocities, along with the proper motions, also allow the orbits of the binaries to be calculated. The orbits found by \citet{Quinn:2009zg} extend to radii $(20-60) \, {\rm kpc}$. In this case the average DM density the binaries experience is significantly, $(50-90)\%$, smaller than the local (i.e.~at the Solar radius) DM density, which further weakens the constraint. \citet{Quinn:2009zg} concluded that the \citet{CG} sample was too small to place meaningful constraints on the halo fraction of compact objects.

\citet{mr} calculated constraints using 251 halo wide binaries from a catalogue compiled by \citet{allen}. 160 of these binaries had radial velocity measurements, allowing their orbits to be calculated. 
Using the binaries which spend the smallest fraction of their time in the Galactic disk, they found that compact objects with $M_{\rm co} \gtrsim 5 \, M_{\odot}$ are excluded from making up all of the halo, and objects with mass $M_{\rm co} \gtrsim 10^{2} M_{\odot}$ make up less than $10\%$, at $95\%$ confidence. Contrary to \citet{Quinn:2009zg}, they found that the average DM densities experienced by the wide binaries are not significantly different from the local density.

In this paper we revisit the modelling assumptions in these analyses, refining several aspects. In particular, previous work assumed that the initial binary semi-major axis distribution is log-flat or a power law, while we use an initial distribution motivated by simulations of the formation of wide binaries during the dissolution of large star clusters~\citep{Kouwenhoven,Daniels_thesis}. We also include unbound binaries in our comparison with observations and take into account the uncertainty in calculating the observed angular separation of a binary from its semi-major axis. We outline our method in Sec.~\ref{sec:method}, present and discuss our results in Sec.~\ref{sec:results}, and conclude with a Summary in Sec.~\ref{sec:summary}.

\section{Method}
\label{sec:method}
\subsection{Binary sample}
\label{sec:sample}
To illustrate the effects of theoretical modelling on the constraints, we use the catalogue of  halo wide binaries compiled from various sources by \citet{allen}. This catalogue was used by \citet{mr} to calculate the most recent wide binary constraints on the abundance of compact objects (that are quoted in reviews of PBH DM e.g.~\citet{Carr:2020xqk,Green:2020jor}).

As discussed by, e.g., ~\citet{CG}, constructing a reliable large catalogue of halo binaries, without selection biases, is non-trivial. Halo binaries need to be distinguished from disk binaries and, as emphasised by \citet{Quinn:2009zg}, radial velocity measurements are required to eliminate chance associations.
\citet{2018MNRAS.480.4302C} constructed a catalogue of halo binaries using Sloan Digital Sky Survey data, however this sample only covers projected separations less than $\sim 0.1 \, {\rm pc}$.

{\em GAIA} \citep{GAIA_DR2} offers the possibility of constructing a large, consistent catalog of halo wide binaries. However at this time there is no definitive sample of halo binaries \citep[see e.g.][for work in this direction]{2017AJ....153..257O,2017AJ....153..259O,2020ApJS..246....4T}.

\subsection{Simulations}
\label{sec:sim}

\subsubsection{Interactions between perturbers and wide binaries}

Our simulations of interactions between perturbers\footnote{We are specifically interested in the case of PBH DM, however the constraints apply to any compact object DM, and therefore we use these terms, and `perturber' interchangably.} and wide binaries largely follow \citet{Yoo:2003fr}.
We assume that all binaries are composed of stars which each have mass $0.5 M_{\odot}$ and that the distribution of the relative velocities of the binaries and perturbers, $f(v_{\rm rel})$, is Maxwellian with dispersion $\sigma_{\rm rel} = 220 \, {\rm km} \, {\rm s}^{-1}$.

When we compare simulated binary distributions with observations in Sec.~\ref{sec:obs} below, the initial binary semi-major axis distribution is taken into account using a scattering matrix formalism, as in \citet{Yoo:2003fr}. In our initial simulations, for simplicity and following previous work, we use a semi-major distribution which is log-flat between $10 \, {\rm au}$ and $10^{5.5} \, {\rm au}$, and assume that the square of the initial eccentricity is uniformly distributed between $0$ and $1$ (i.e. thermal).

As in previous work~\citep{Yoo:2003fr,Quinn:2009zg,mr} we do not include perturbations from Giant Molecular Clouds (GMCs) or the effects of Galactic tides. Due to their low  number density in the halo the impact of GMCs on halo wide binaries is expected to be small, and neglecting it is a conservative assumption. Galactic tides are smaller for halo wide binaries than for the disk binaries studied in \citet{j+t}, and likewise including their effects would act to tighten the constraints. We have also assumed that the PBHs are smoothly distributed and are not themselves in binaries. Some PBHs are expected to form binaries in the early Universe~\citep{Nakamura:1997sm,Ali-Haimoud:2017rtz}, and PBH clusters form not long after matter-radiation equality~\citep{Afshordi:2003zb,Inman:2019wvr}. The evolution of these clusters, and in particular the disruption of PBH binaries within them, is a challenging problem and the present day spatial distribution of PBHs within galaxies is not yet understood in detail. 

\begin{figure*}
   \centering
  \includegraphics[width=0.8\textwidth]{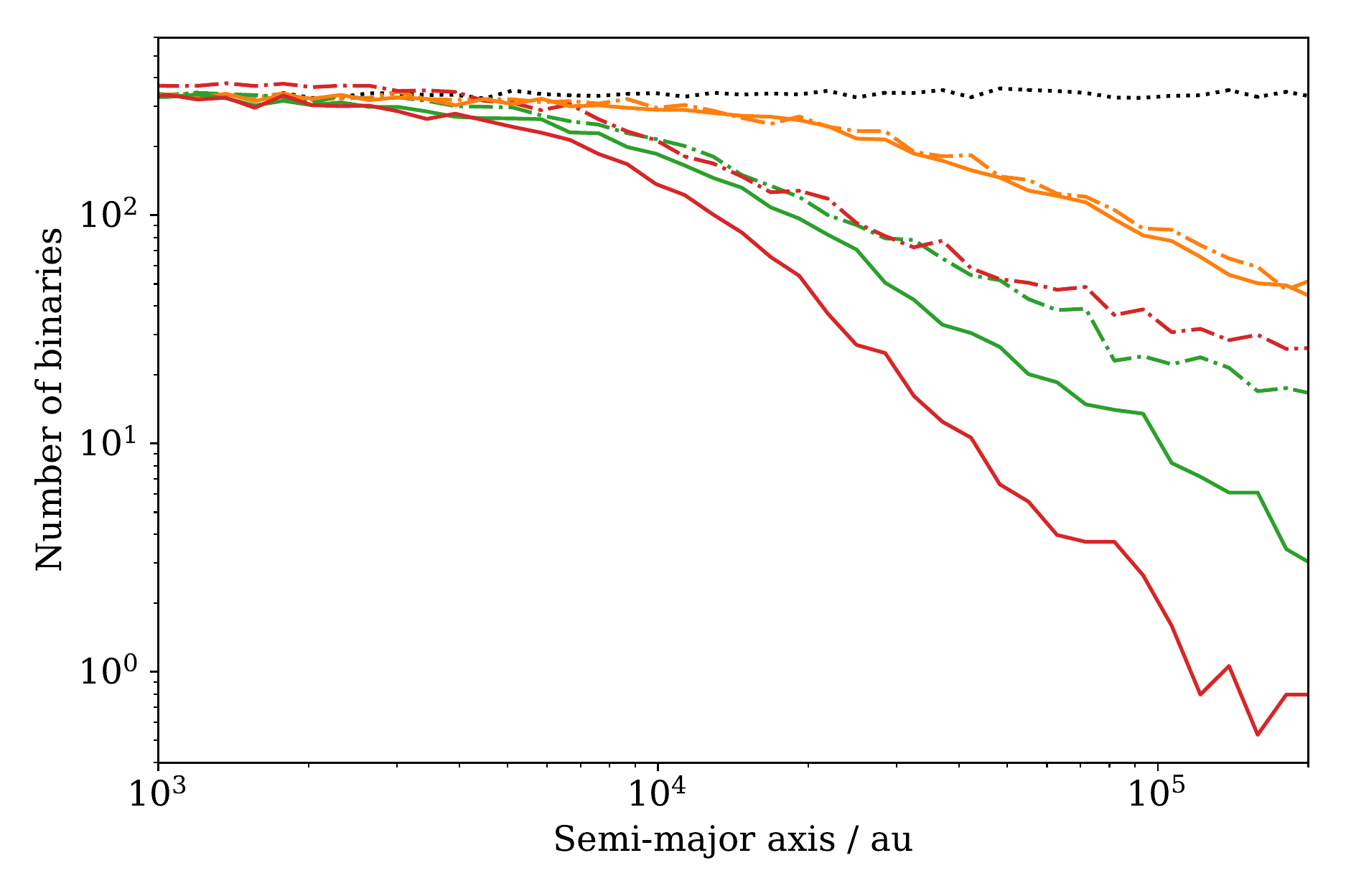}
   \caption{The final semi-major axis distribution of $10^5$ binaries composed of stars with mass $0.5 M_\odot$ evolved for 10 Gyr in a population of perturbers with a Maxwellian relative velocity distribution with dispersion $\sigma_{\rm rel} = 220 \, {\rm km \, s}^{-1}$, mass density $\rho= 0.009M_\odot$\, pc$^{-3}$ and masses 10$M_\odot$ (orange lines), 100$M_\odot$ (green) and 1000$M_\odot$ (red). The dot-dashed lines are for the full binary population (bound and unbound binaries), while the solid lines show only the binaries that remain bound at all times. The initial, log-flat, binary semi-major axis distribution is shown by the black dotted line.
  \label{fig:sim} }
    \end{figure*}
 
Unlike previous work on constraints on compact object DM from halo binaries, we include unbound binaries in our comparison with observed binaries. \citet{Yoo:2003fr} argued that disrupted binaries quickly diffuse to large separations, beyond those probed observationally. However \citet{j+t} included unbound systems in their study of the effects of perturbers on disk binaries using diffusion equations.  They found that the stars from unbound binaries have small relative velocities, which would lead them to be detected as binaries by surveys. Furthermore, they also found that some unbound binaries can become rebound.

The rate at which encounters with impact parameter between $b$ and $b + {\rm d} b$ and relative velocity between $v_{\rm rel}+ {\rm d} v_{\rm rel}$ occur, $\dot{C}$, is given by
\begin{equation}
\label{cdot}
    \dot{C} = n_{\rm p} v_{\rm rel} 2 \pi b \, {\rm d} b f(v_{\rm rel}) \, {\rm d} v_{\rm rel} \,,
\end{equation}
where $n_{\rm p}= \rho/M_{\rm p}$ is the perturber number density and $\rho$
and $M_{\rm p}$ are the perturber mass density and mass respectively. We consider perturber masses in the range $1 M_{\odot} < M_{\rm p} < 3 \times 10^{3} M_{\odot}$ and fix $\rho$ to the standard value for the local DM density, $0.009M_\odot$pc$^{-3}$ (e.g. \citet{deSalas:2020hbh}), however the constraints can be straight forwardly rescaled to other values of the local DM density.

We have found (see Fig.~3.5 of \citet{Emily_thesis}) that encounters which cause a fractional change in the binary energy less than $0.1\%$ have a negligible (less than $0.1\%$)
effect on the semi-major axis distribution, therefore we do not include these encounters in our simulations.
We calculate the number of interactions expected within a time $T=10 \, {\rm Gyr}$, roughly equally to the age of the MW. For each individual binary the actual number of encounters experienced is drawn from a Poisson distribution and the impact parameter and relative velocity of each encounter are found from the distributions in Eq.~(\ref{cdot}).

The relative velocity between the perturber and binary is always much larger than the orbital velocities of the binary stars. Therefore the stars can be treated as stationary during an encounter and the impulse approximation used to calculate its effect (e.g. \citet{b+t}). The positions of the stars are unperturbed, while the changes in their velocities are perpendicular to the trajectory of the perturber and given by
\begin{equation}
\Delta v_{i} = \frac{2 G M_{\rm p}}{v_{\rm rel} b_{i}} \frac{{{\bf b}_{i}}}{b_{i}} \,,
\end{equation}
where ${\bf b}_{i}$ is the impact parameter to star $i$.

Binaries are evolved in time between encounters. For bound binaries the time between encounters is much longer than the period of the binary, so we do this by taking a random value for the mean anomaly between $0$ and $2 \pi$ and converting this (via Kepler's equation) to a future true anomaly. The hyperbolic orbits of unbound binaries are not periodic, so in this case we evolve the binary's eccentric anomaly forwards in time exactly. The position and velocity vectors of the two stars before each encounter are calculated from their semi-major axis, eccentricity and orbital phase (true anomaly). 

Fig.~\ref{fig:sim} shows the final semi-major axis distribution for simulations with a log-flat initial binary semi-major axis distribution and perturbers with density $\rho= 0.009 \, M_{\odot} \, {\rm pc}^{-3}$ and masses $M_{\rm p} = 10, 10^{2}$ and $10^{3} M_{\odot}$. It shows both the full binary population (dot-dashed lines) and also just the binaries which remain bound throughout the whole simulation (solid lines), i.e. the result that would be obtained by discarding unbound binaries.
We see that for $M_{\rm p} = 10^{2}$ and $10^{3} M_{\odot}$ (green and red lines respectively) the two distribution differ significantly for $a \gtrsim 10^{4} \, {\rm au}$, and hence discarding unbound binaries significantly underestimates the abundance of the widest observed apparent binaries. As mentioned previously, \citet{j+t} find that disrupted binaries in the Galactic disk have very small relative velocities. For perturbers larger than $\sim 1 M_{\odot}$, however, the increase in relative velocity due to encounters is more significant \citep[Eq.~A2]{Yoo:2003fr}. \color{black} We note that our results for binaries which remain bound throughout are in good agreement with previous work by \citet{Yoo:2003fr} and \citet{mr}.

 The large abundance of unbound wide binaries for $M_{\rm p} = 10^{3} M_{\odot}$ is likely due to the low number density of perturbers, which decreases with increasing perturber mass (for constant perturber mass density). Even though encounters with $M_{\rm p} = 10^{3} M_{\odot}$ are more likely to break the binaries, multiple encounters are required to give the binaries sufficient relative velocity to drift apart within the timescale of the simulation. This may also explain why for $M_{\rm p} = 10 M_{\odot}$ there are very few unbound binaries; these binaries have experienced a large number of encounters giving them sufficient relative velocity to drift far apart by the end of the simulation.\color{black}

\subsubsection{Orbits of binaries}
\label{sec:orbit}

\begin{figure*}
   \centering
  \includegraphics[width=0.8\textwidth]{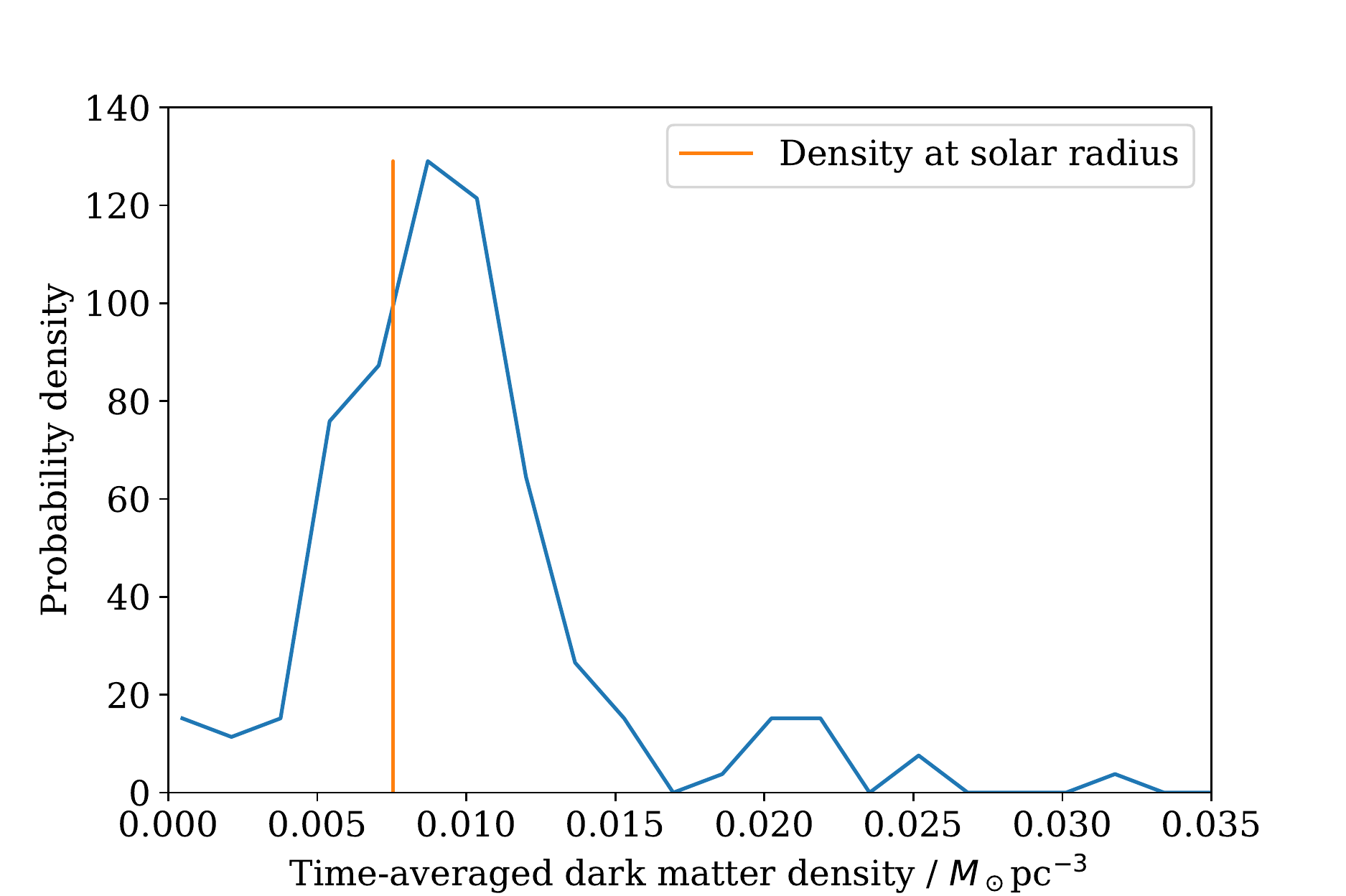}
   \caption{The probability distribution of the time-averaged dark matter density calculated along the orbits of 160 binaries from \citet{allen} that it is possible to calculate orbits. The orange vertical line shows the dark matter density at the Solar radius, 0.00754 $M_\odot$ pc$^{-3}$. 
  \label{fig:galpy_avg_dm_density_dist} }
    \end{figure*}

It is useful to calculate the orbits of the wide binaries within the MW potential for two reasons. Firstly each binary experiences an orbit-dependent, time-varying, DM density. This can be taken into account by finding the time-averaged DM density along each binary orbit, and 
scaling the constraint on the perturber density by the mean time-averaged DM density divided by the value of the local DM density \citep{Quinn:2009zg}. Secondly, binaries will experience perturbations from stars when passing through the Galactic disk, and hence binaries which spend the smallest fraction of their orbits within the Galactic disc are more powerful for constraining perturbers in the halo. \citet{mr} classified the binaries as `most halo-like' according to the fraction of time their orbit spends within the disc ($|z| < 500 \, {\rm pc}$). 

We calculated the binary orbits for the 160 binaries in the \citet{allen} catalog~\footnote{Online data from \href{https://cdsarc.cds.unistra.fr/viz-bin/cat/J/ApJ/790/158}{https://cdsarc.cds.unistra.fr/viz-bin/cat/J/ApJ/790/158}.} which have sufficient data to do this using the $\texttt{galpy}$ Python package \citep{Bovy}. For each binary we use the most recent data from the SIMBAD database~\citep{2000A&AS..143....9W}, usually from GAIA DR2~\citep{GAIA_DR2}.
We used the $\texttt{MWPotential2014}$ model in $\texttt{galpy}$, which has a Navarro-Frenk-White density profile~\citep{Navarro:1996gj} for the MW halo, along with potentials for the disk and bulge. While this model is not intended to be the best current model of the MW, its parameters are similar to those obtained from, e.g., fits to rotation curve data~\citep{recent_galactic_model}, and it is sufficiently accurate for our purpose.
We find the mean time-averaged DM density for the 160 binaries is 
$\sim 40\%$ larger than the DM density at the solar radius.
\citet{Quinn:2009zg} et al. found substantially smaller time-averaged DM densities for the widest binaries that they studied. However, like \citet{mr}, we find that the orbit for NLTT10536 reaches a maximum $z$ value of around $5 \, {\rm kpc}$, whereas the orbit calculated by \citet{Quinn:2009zg} extended to $z \approx 40 \, {\rm kpc}$. Also, using the most recent determination of its distance, proper motion and radial velocity, we find an orbit for NLTT16394 which is confined to smaller values of $z$ and $R$ than previously found~\citep{mr,Quinn:2009zg} 

The probability density of the time-averaged dark matter densities for the 160 binaries it is possible to calculate orbits for is shown in Fig.~\ref{fig:galpy_avg_dm_density_dist}.
 The distribution of time-averaged DM densities experienced by the binaries is not too wide (full width at half maximum $0.007 \,M_\odot \, $pc$^{-3}$). This suggests that simply scaling the constraint on the perturber density by the mean time-averaged DM density should capture the effect of the varying DM density experienced by the binaries.

\begin{figure*}
   \centering
  \includegraphics[width=0.8\textwidth]{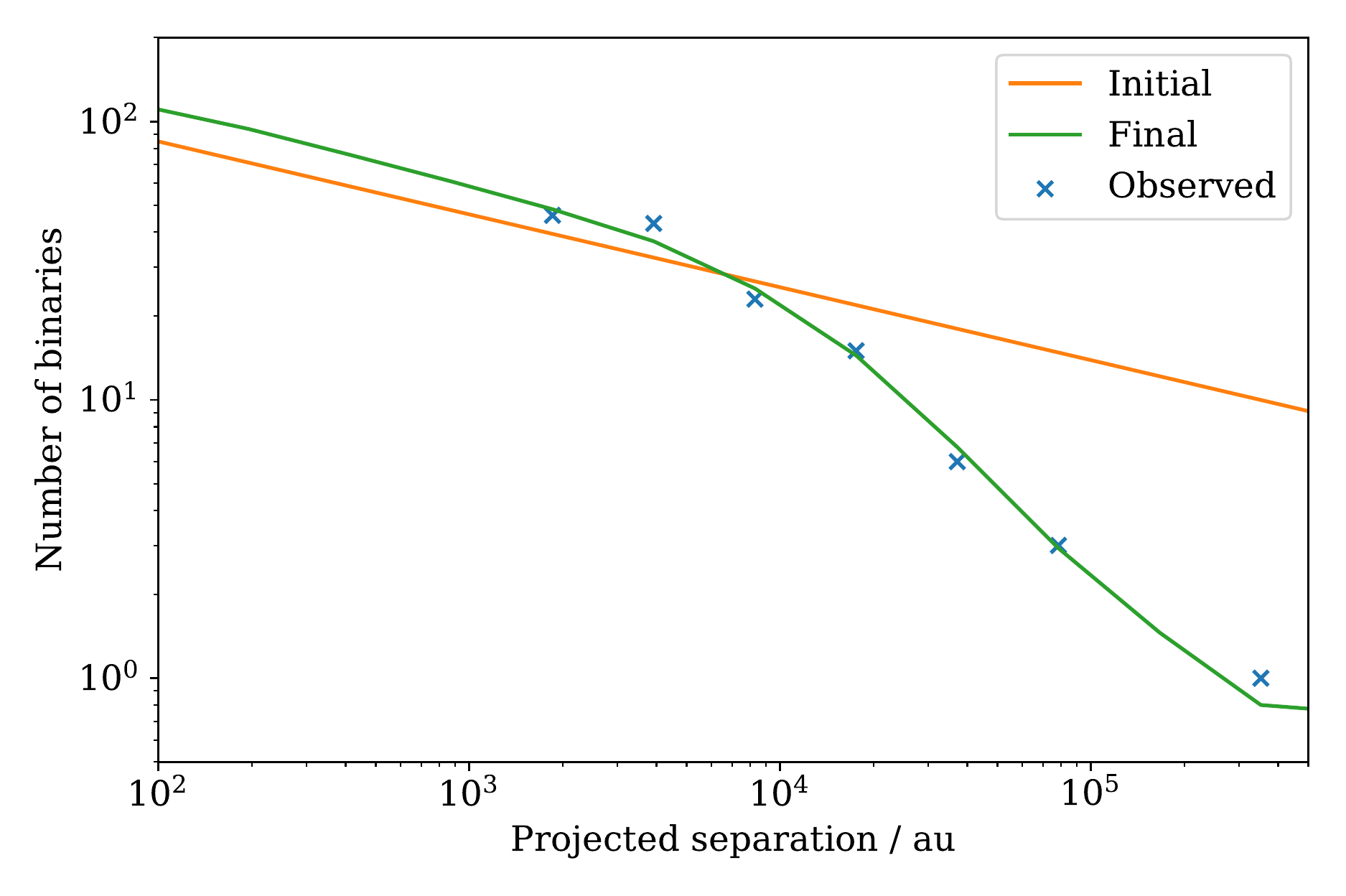}
  \caption[The best fit final projected separation distribution]{The best fit final projected separation distribution (green line) compared with the observed separation distribution (blue crosses). The corresponding initial distribution (orange line), which has parameters $\alpha=1.26$ and $A=1.00$ is also shown. The best fit perturber mass and density are $\Mp=30\, M_\odot$ and $\rho=0.012 \, M_\odot\, $pc$^{-3}$ respectively.}
    \label{fig:best_fit_dist} 
 
    \end{figure*}

\subsection{Comparison with observations}
\label{sec:obs}

\subsubsection{Initial semi-major axis distribution}

A model is required for the initial semi-major axis separation distribution from which the current distribution has evolved.  Unfortunately, it is extremely unclear what that initial distribution should be. 
Previous work on wide binary disruption \citep{Yoo:2003fr,Quinn:2009zg,mr,WSW,j+t} used a power law distribution, $\propto a^{-\alpha}$, which is the simplest generalisation of \"{O}pik's Law, a log-flat distribution. It is not at all obvious that this simple distribution is a good model for the initial wide binary semi-major axis distribution  \citep[see also][]{2020ApJS..246....4T}.

Binary semi-major axis distributions usually seem to follow a roughly log-normal distribution with a peak at tens to hundreds of au depending on the primary mass \citep[see e.g.][]{2010ApJS..190....1R,Duchene2013,2015MNRAS.449.2618W}.
The best understood sample of binary separations are local field G dwarfs \citep{2010ApJS..190....1R} which have a log-normal separation distribution which peaks at $\sim 30$ au, with a variance of 1.5 in the log (so roughly two thirds of systems lie between 1 and 1000 au).  


Local field G dwarfs have a few per cent of very wide binaries beyond $10^4$ au, which is usually modelled as the exponential tail of the G dwarf log-normal.  However, it is not clear that this is a good way of modelling the wide binary tail. The formation mechanism(s) of very wide binaries, with semi-major axis $>10^4$ au, are not  understood.  The peaks of binary distributions (at tens to hundreds of au) are thought to arise from core and/or disc fragmentation during star formation  \citep[see][]{2007prpl.conf..133G,Duchene2013,2014prpl.conf..267R}. However, systems with separations $>10^4$ au are much wider than the size of star forming cores and so it is uncertain how they arise.  The most likely mechanism suggested so far is `soft capture' \citep{Kouwenhoven,2010MNRAS.404..721M,2011MNRAS.415.1179M}, where a wide binary is formed by the chance proximity of two stars with low relative velocities during the dissolution of a star cluster or star forming region.

\begin{figure*}
   \centering
  \includegraphics[width=0.8\textwidth]{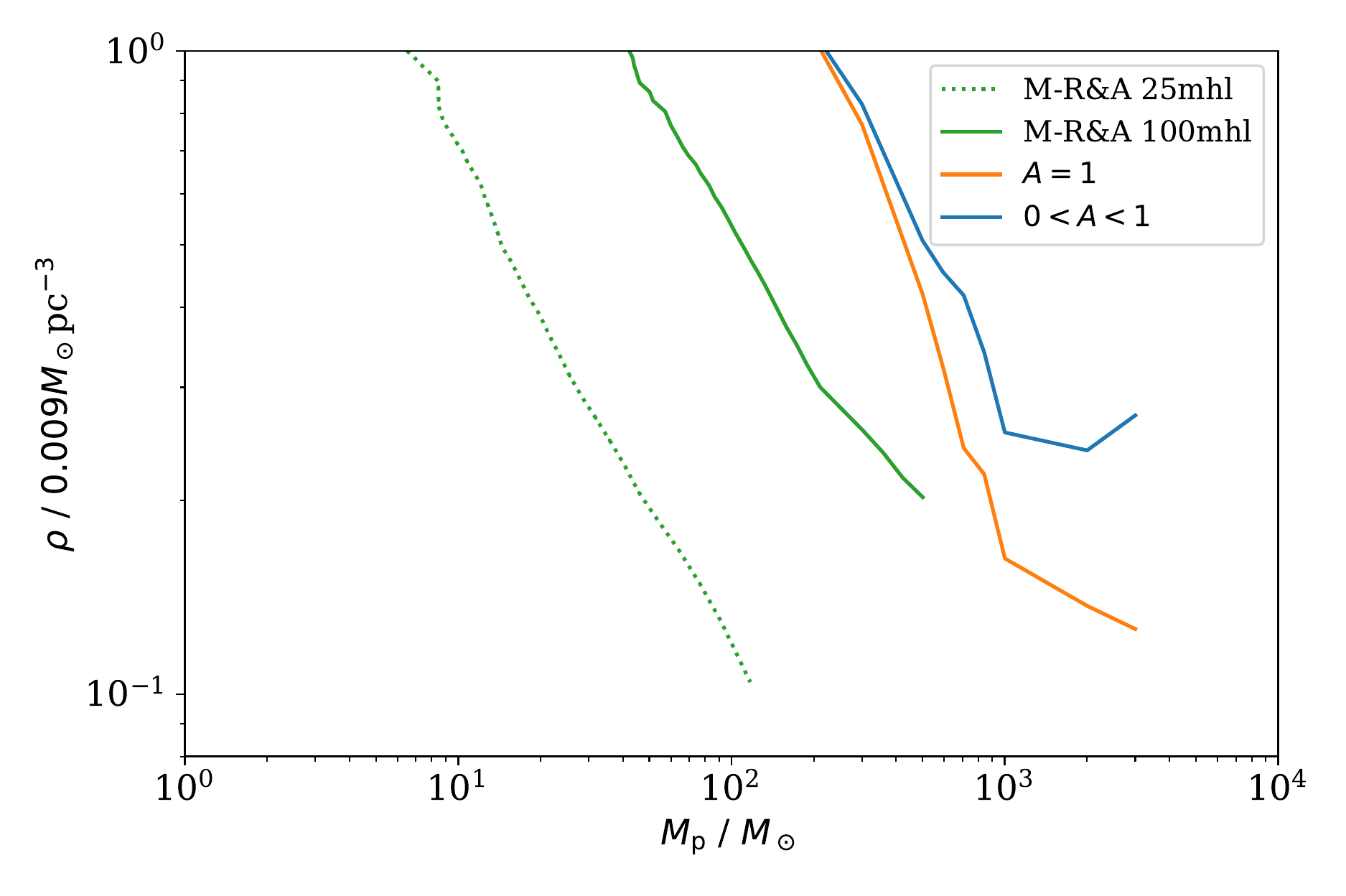}
   \caption{Two sigma constraints on the perturber density, $\rho$, as a function of the perturber mass, $M_{\rm p}$. The orange and blue lines show our constraints for
   $A=1$ (initial binary semi-major axis distribution is a pure power law) and $0<A<1$ (allowing a varying fraction of the initial distribution to be log-normal) respectively. The dotted and solid green lines are the \citet{mr}
   constraints for their 25 and 100 most halo like binaries respectively.
  \label{fig:results} }
    \end{figure*}

Simulations of soft capture show that the rate is low, but that very wide binaries can be formed. \citet{Daniels_thesis} carried out simulations of the dissolution of clusters with different levels of (fractal) substructure in the initial star cluster \citep[c.f.][]{Kouwenhoven}. From their simulations we find that a power law distribution is a good fit to the wide binaries formed via soft capture (see e.g. their Fig.~5.7), with the slope decreasing from $\alpha = 0.9$ to $0.7$ as the level of substructure decreases.  This could well appear like an exponential tail in the broader distribution of separations (as current data is too poor to show any features of different formation mechanisms).

How many wide binaries we would expect is another unknown.  The fraction of wide binaries in the local field G dwarf population is a few per cent (depending on exactly where one draws the line for wide binaries, see e.g. \citet[][]{2012AJ....144..102T}).  However, the local field population should have been processed to some degree by other field stars in exactly the same way a PBH population would process the halo binaries.  Therefore this provides a lower limit on wide binary production in what are now Galactic disc field stars.  If we assume soft capture as the mechanism then we would not expect a metallicity-dependence on the primordial wide binary fraction\footnote{ \citet{el-badry19} find a very slight excess of metal rich field wide, ($5 \, 000-50 \, 000$) au, binary systems over metal poor systems, but the two are very similar.}.

Therefore as well as considering a pure power law for the initial binary semi-major axis distribution (motivated by our fits to simulations of soft capture), we also study an initial distribution where in addition primordial binaries make up a variable fraction, $1-A$, of the total population between $a_\mathrm{min}=30$ au and $a_\mathrm{max}=2 \times10^4$ au. We assume that the primordial binaries have a log-normal distribution with mean $\mu=100$ au and log width $\sigma=1.5$ (which is closer to the local pre-main sequence binary population than the local field, see \citet{Duchene2013}).

\subsubsection{Binary separations}

The observed separation of a system is the angular separation, which depends on its semi-major axis, eccentricity, phase, inclination, orientation, and distance. From a single observation of a separation on the sky it is impossible to determine the true semi-major axis in anything other than a purely statistical way. \citet{Yoo:2003fr} calculated a theoretical angular separation distribution by convolving the projected separation distribution of their simulated binaries with their assumed (inverse) distance distribution. \citet{mr} instead compared the semi-major axis distribution of simulated and observed binaries, using a statistical relationship between semi-major axis and angular separation to estimate the observed semi-major axes. 

The problem with using a statistical relationship between the instantaneous separation and the semi-major axis is that it only holds for a `typical' binary. On average, the semi-major axis of a binary is slightly larger than the observed separation (how much larger depends on the assumed eccentricity distribution).  However, some binaries (high eccentricity systems at apastron, oriented such that we see the 3D separation in 2D) will be observed with a separation of approximately twice the semi-major axis.  Such systems are rare, but will tend to fall at the widest extreme of the distribution.  Therefore, at the widest end of the distribution this would tend to over-estimate the semi-major axes. For this reason we compare the projected separations of our theoretical distribution with the observed distribution, by randomising the viewing angles, rather than attempting to turn the observed separation distribution into a semi-major axis distribution.

To calculate the predicted separation distribution for a given initial semi-major axis distribution, we use the same scattering matrix formalism as \citet{Yoo:2003fr}. Since each binary evolves independently, then
the expected number of binaries with projected separation $r_j$, $P(r_j, \Mp,\rho)$, is given by
\begin{equation}
    P(r_j, \Mp,\rho) \propto a_{j} \, S_{ij}(\Mp,\rho) \, q(a_j),
\end{equation}
where $q(a)$, is the probability density of the initial semi-major axis distribution and
the scattering matrix, $S_{ij}(\Mp,\rho)$, is the number of simulated binaries with initial semi-major axis in the $i$-th logarithmically spaced bin centered at $a_i$ that have final projected separation $r_j$ for a simulation with perturber mass $\Mp$ and dark matter density $\rho$. The factor of $a_j$ appears because our semi-major axis bins are logarithmically spaced.

\subsubsection{Statistical analysis}

Previous work has used likelihood analysis \citep{Yoo:2003fr} or the Kolmogorov-Smirnov (K-S) test \citep{mr} to compare simulated and observed binary distributions. Both of these methods have drawbacks for this analysis. Likelihood analysis doesn't provide information about how good a fit the best fit is, while the K-S test is less sensitive to differences in the extremes of distributions, which is suboptimal as the widest binaries are most affected by perturbers. The classical $\chi^2$ test is not valid if the number of samples in any bin is small, which is the case for the widest binaries. We therefore use a modified version of the $\chi^2$ test, which provides $p$-values, is valid for small sample sizes, and is equally sensitive to deviations across the whole range of the distributions.

The modified $Y^2$ statistic \citep{Lucy}, is rescaled so that its
variance is fixed to be equal to twice its mean, and hence the standard translation of $\chi^2$ values into $p$-values is valid, even for small samples. The $Y^2$ statistic is defined as
\begin{equation}
    Y^2 = \nu + \sqrt{\frac{2\nu}{2\nu + \Sigma_i n_i^{-1}}}\left(\chi^2-\nu\right) \,,
\end{equation}
where $n_i$ is the expected number of binaries in the $i$-th bin. The number of degrees of freedom, $\nu$, is equal to the number of bins minus the number of fitted parameters plus one as the $n_i$'s have been normalised to match the total observed number of binaries. The $\chi^2$ statistic is given, as usual, by
\begin{equation}
    \chi^2 = \sum_i \frac{(N_i-n_i)^2}{n_i} \,,
\end{equation}
where $N_i$ is the number of observed binaries in the $i$-th bin and the sum is over all bins with non-zero $N_i$.

\section{Results and discussion}  
 \label{sec:results}

We calculate the $Y^2$ statistic as a function of perturber mass, $\Mp$, and density, $\rho$, the fraction of the binaries that have power law semi-major axis distribution initially, $A$, and the slope of the power law, $\alpha$. For each $\Mp$ and $\rho$ combination we find the minimum value of $Y^2$, $Y^2_\text{min}(\Mp,\rho)$. We first check that the best fit is a sufficiently good fit by comparing the global minimum value of $Y^2$,   $Y^2_\text{min}$, with the number of degrees of freedom, $\nu$. Here we have two fitted parameters ($A$ and $\alpha$) and seven bins, so $\nu = 7- (2+1)=4$. The global best fit has $\alpha=1.26$, $A=1$, $M_{\rm p}=30 M_{\odot}$ and $\rho= 0.012 M_{\odot} \, {\rm pc}^{-3}$. It has $Y^2_{\rm min} < 3$ and hence is indeed a good fit to the data. Fig.~\ref{fig:best_fit_dist} compares the best fit projected separation distribution with the observed separation distribution, and also shows the corresponding initial separation distribution.

Next we calculate constraints on $\Mp$ and $\rho$ by finding the pairs of values for which 
\begin{equation}
    \Delta Y^2(\Mp,\rho)=Y^2_\text{min}(\Mp,\rho)-Y^2_\text{min} =  \text{inverse}\left(1-\text{cdf}(p)\right),
\end{equation}
where $p= 0.05$ for $2\sigma$ constraints, and $\text{cdf}$ is the cumulative distribution function of the $\chi^2$ distribution with 2 degrees of freedom, since we are now finding constraints on two parameters ($\Mp$ and $\rho$). 
We do this for both $A=1$, i.e.~a pure power law distribution for the initial binary distribution, and $0<A<1$, i.e.~allowing a varying fraction of the distribution to be log-normal. Finally, as discussed in Sec.~\ref{sec:orbit}, we rescale our constraints by a factor of $0.71$ to take into account the average DM density experienced by the binaries along their orbits.

Our constraints on the perturber mass, $\Mp$, and density, $\rho$, are shown in Fig.~\ref{fig:results}. We compare our (very similar) $2 \sigma$ constraints for $A=1$ (orange line) and $0<A<1$ (blue line) with the \citet{mr} constraints from their 100 and 25 `most halo like' binary samples (green solid and dashed lines respectively). For values of $\Mp$ larger than those plotted, the \citet{mr} constraints are expected to be roughly constant.

We tested the validity of comparing 25 observed binaries with our simulations and found that randomly choosing groups of 25 binaries resulted in constraints that varied significantly. This is due to the large stochasticity in the distribution of observed angular separations from a semi-major axis distribution when the number of binaries is small.  This suggests that a much larger sample of halo wide binaries is required to provide any meaningful constraints.  Therefore we only present our constraints
calculated using the full sample of binaries to avoid this stochasticity. Fig.~7 of \citet{mr} indicates that they were able to calculate reliable constraints from small sub-populations of binaries. This difference is likely to be because they compare `virtual' binaries, constructed from $500-10,000$ simulated binaries, with the semi-major axis of observed binaries calculated by assuming there is a one-to-one relationship between projected separation and semi-major axis. This assumption is an oversimplification that does not take into account the varied phases and orientations of the observed binaries.

Our constraint is significantly weaker than that from \citet{mr}. We find $f_{\rm co} <  1$ for $M_{\rm p} \approx 300 \, M_{\odot}$, tightening with increasing $M_{\rm p}$ to $f_{\rm co} < 0.26$ for  $M_{\rm p} \gtrsim 1000 \, M_{\odot}$. An obvious question is `why are our constraints so much weaker than those of \citet{mr}?'.
To restate the obvious - compact objects destroy wide binaries, and the wider the binary, the more susceptible to destruction it is.  Therefore, the constraints on the allowed compact object density are extremely sensitive to the
number of very wide binaries, and the exact values of the semi-major axes. We include two effects that \citet{mr} did not, both of which act to increase the number of very wide binaries predicted for any particular initial semi-major axis distribution and perturber population. Consequently, the abundance of perturbers required to reduce the abundance of the widest binaries below that which we observe is larger.

Firstly, we do not discard unbound binaries.  This means there are systems with wide separations which, from a single observation, would be indistinguishable from a (very weakly) bound `true' binary.  This increases the number of very wide systems that could potentially be observed. 

Secondly, by projecting our theoretical distribution into observed separations we correctly allow for systems to be observed where the separation is significantly larger than the semi-major axis (up to a factor of two for bound binaries, and greater than two for unbound systems).  Such systems are rare, but by definition fall at the widest extreme of the distribution which is what sets the constraints.

The inclusion of unbound binaries in the final distribution contributes the most to weakening the constraints. Fig.~\ref{fig:sim} shows that at the largest semi-major axis, the total number of binaries is at least one magnitude larger than the number of bound binaries for $\Mp > 100 M_{\odot}$. The next largest contribution is from the initial semi-major axis distribution. For perturber masses $\Mp > 1000 M_{\odot}$, the fraction of dark matter that could consist of compact objects (Fig.~\ref{fig:results}) increases from ~0.1 to ~0.3 when comparing a variable distribution ($0<A<1$) with a power law distribution ($A=1$). Comparing projected separations, and therefore taking into account the large apastron distance of wide binaries, is likely to have had a relatively small effect on the final constraints. While the number of binaries at the largest separations, which are most susceptible to this effect, are the most important for calculating constraints, the increase in binary separation due to this effect is approximately a factor of 2 in most cases. \color{black}

\section{Summary}  
 \label{sec:summary}

 We have revisited the theoretical modelling involved in placing constraints on the fraction of the MW halo in compact objects from the dynamical effects on the semi-major axis distribution of wide binary stars. We have improved on previous work in several ways. We have used a physically motivated model for the initial binary semi-major axis, taken into account the uncertainty in relating semi-major axis to observed angular separation, and retained unbound binaries. We compare simulated binary separations with observations using the $Y^2$ statistic  \citep{Lucy}. This retains the advantages of the $\chi^2$ statistic, namely it allows the goodness of fit of the best fit to be checked and (unlike the K-S test) is sensitive to deviations at the extremes of the distributions.

We find that with these improvements the constraints obtained using the \citet{allen} wide binary sample are significantly weakened. We find $f_{\rm co} <  1$ for $M_{\rm co} \approx 300 \, M_{\odot}$, tightening with increasing 
$M_{\rm co}$ to $f_{\rm co} < 0.26 $ for  $M_{\rm co} \gtrsim 1000 \, M_{\odot}$, whereas \citet{mr} found $f_{\rm co} < 1$ for  $M_{\rm p} \sim 10 \, M_{\odot}$, tightening with increasing $M_{\rm co}$ to $f_{\rm co} < 0.1$ for  $M_{\rm co} \gtrsim 100 \, M_{\odot}$. It is therefore crucial that these modelling improvements are implemented when calculating constraints on compact objects using future improved catalogs of halo wide-binaries. 


\section*{Acknowledgements}

ET was supported by a United Kingdom Science and Technology Facilities Council (STFC) studentship. 
  AMG is supported by STFC grant ST/P000703/1.  
  For the purpose of open access, the authors have applied a CC BY public copyright licence to any Author Accepted Manuscript version arising. This research has made use of the SIMBAD database, operated at CDS, Strasbourg, France. \\

\section*{Data Availability}

 This work is entirely theoretical, and has no associated data.



\bibliographystyle{mnras}
\bibliography{wb} 








\bsp	
\label{lastpage}
\end{document}